\documentclass[twoside,11pt]{article}

\usepackage{obs_study_style}
\usepackage{amsmath}
\usepackage{multirow}
\usepackage{booktabs}
\usepackage{tabularx}
\usepackage{array}
\usepackage{rotating}
\usepackage{float}
\usepackage{xcolor}
\usepackage[normalem]{ulem}
\useunder{\uline}{\ul}{}
\usepackage{adjustbox}
\usepackage{makecell}
\usepackage{enumitem}
\usepackage{rotating}
\usepackage{ragged2e}
\usepackage{hyperref}

\newcolumntype{C}[1]{>{\centering\arraybackslash}p{#1}}

\heading{}{}{}{}{}{Jiahao Zhang and Bojian Feng and Qingyan Xiang}

\ShortHeadings{LLMs for sensitivity analysis}{Zhang, Feng, and Xiang}
\firstpageno{1}

\begin{document}

\title{Using large language models for sensitivity analysis in causal inference: case studies on Cornfield inequality and E-value}

\author{\name
       \name Qingyan Xiang \email qingyan.xiang@vumc.org \\
       \addr Department of Biostatistics\\
       Vanderbilt University Medical Center\\
       Nashville, TN, USA
       \AND
      \name Jiahao Zhang\email zhangjh325@mail2.sysu.edu.cn \\
       \addr Department of Mathematics\\
       Sun Yat-sen University\\
       Guangzhou, Guangdong province, China 
       \AND
        \name Bojian Feng \email fengbj@zjcc.org.cn \\
       \addr Department of Diagnostic Ultrasound Imaging \& Interventional Therapy\\
       Zhejiang Cancer Hospital\\
       Hangzhou, Zhejiang province, China
}

\maketitle

\begin{abstract}

Sensitivity analysis methods such as the Cornfield inequality and the E-value were developed to assess the robustness of observed associations against unmeasured confounding -- a major challenge in observational studies. However, the calculation and interpretation of these methods can be difficult for clinicians and interdisciplinary researchers. Recent advances in large language models (LLMs) offer accessible tools that could assist sensitivity analyses, but their reliability in this context has not been studied. We assess four widely used LLMs, ChatGPT, Claude, DeepSeek, and Gemini, on their ability to conduct sensitivity analyses using Cornfield inequalities and E-values. We first extract study-specific information (exposures, outcomes, measured confounders, and effect estimates) from four published observational studies in different fields. Using such information, we develop structured prompts to assess the performance of the LLMs in three aspects: (1) accuracy of E-value calculation, (2) qualitative interpretation of robustness to unmeasured confounding, and (3) suggestion of possible unmeasured confounders.  To our knowledge, there has been little prior work on using LLMs for sensitivity analysis, and this study is an early investigation in this area. The results show that ChatGPT, Claude, and Gemini accurately reproduce the E-values, whereas DeepSeek shows small biases. Qualitative conclusions from all the LLMs align with the magnitude of the E-values and the reported effect sizes, and all models identify biologically and epidemiologically plausible unmeasured confounders. These findings suggest that, when guided by structured prompts, LLMs can effectively assist in evaluating unmeasured confounding, and thereby can support study design and decision-making in observational studies.

\end{abstract}


\section{Introduction}


In observational studies, the interest is often to estimate the causal effect of exposure on  outcome, which requires appropriate causal inference methods. These methods typically rely on strong assumptions, most notably the no unmeasured confounding assumption \citep{brumback2004sensitivity, hernan2010causal}. However, in many studies, even after carefully controlling for measured covariates, unmeasured confounding may still exist, which can lead to biased effect estimates and misleading  conclusions \citep{vanderweele2011unmeasured, zhang2020assessing, gaster2023quantifying}.

To assess the impact of unmeasured confounders on the effect estimates in observational studies, many sensitivity analysis methods have been developed, including the Cornfield inequality  and the E-value. The Cornfield inequality was initially proposed to address the controversies around smoking and lung cancer \citep{cornfield1959smoking}, which introduced one of the earliest approaches for sensitivity analysis. Related to their concept, \citet{vanderweele2017sensitivity} proposed the E-value, a metric representing the minimum strength of the joint association of unmeasured confounders with exposure and outcome to fully explain away the observed association. The E-value is now widely used in observational studies for sensitivity analyses for unmeasured confounding \citep{blum2020use}.

\par

For clinicians and interdisciplinary researchers who are conducting observational studies, it can be challenging to fully understand, calculate, and interpret the Cornfield inequality and the E-value. However, advances in large language models (LLMs) offer accessible tools that could assist with sensitivity analyses. LLMs represent a class of artificial intelligence models designed to understand and generate human-like text \citep{vaswani2017attention, brown2020language, zhao2023survey}. With proper input information from a study, we hypothesize that LLMs can support the calculations of the  Cornfield inequality and the E-value, generate clear explanations, and suggest possible unmeasured confounders.

LLMs have increasingly been explored as tools for causal inference \citep{liu2025large, ma2025causal}. Prior studies have primarily focused on using LLMs for tasks such as   causal discovery \citep{jin2023can, takayama2024integrating, cohrs2024large, lee2025incorporating} and causal reasoning \citep{yao2023tree, jin2023cladder, chi2024unveiling}. However, using LLMs for sensitivity analysis for unmeasured confounding remains largely unexplored, and to the best of our knowledge, this study is an early investigation of this topic.

In this article, we assess the performance of the LLMs in performing sensitivity analysis using case studies from four published observational studies. The studies cover topics related to smoking, back pain, Alzheimer's disease, and environmental health. We will use the information extracted from these studies to evaluate four widely used LLMs. Our assessment is threefold: (i) quantitatively assessing if the LLMs can correctly compute the E-values, (ii) qualitatively assessing if the LLMs can generate proper conclusions on how likely the study findings are subject to unmeasured confounding, and (iii) assessing if the LLMs can suggest potential unmeasured confounders.  Through this evaluation, we hope our prompting strategy can serve as a tool to help researchers better assess whether findings from observational studies are subject to unmeasured confounding, thereby helping decision-making in clinical and public health research.

\section{Methods}

\subsection{Brief overview}



\citet{cornfield1959smoking} discussed the issues against the causal role of tobacco smoking in lung cancer, and their idea was later formalized as the Cornfield inequality. They illustrated this idea with the example of smoking and lung cancer: if smokers have a ninefold higher risk of lung cancer compared to non-smokers, and if this risk is due entirely to some unmeasured factor rather than to smoking itself, then this factor would need to be extremely strongly associated  with smoking (with at least a nine-fold risk ratio). However, since no such factor has been identified, the authors argued that confounding alone could not account for the observed association between smoking and lung cancer.  This reasoning, which evaluates whether an unmeasured confounder could account for an observed effect, later became a core concept in sensitivity analysis for causal inference.
\begin{figure}[tb]
\centering
\includegraphics[width=0.86 \textwidth]{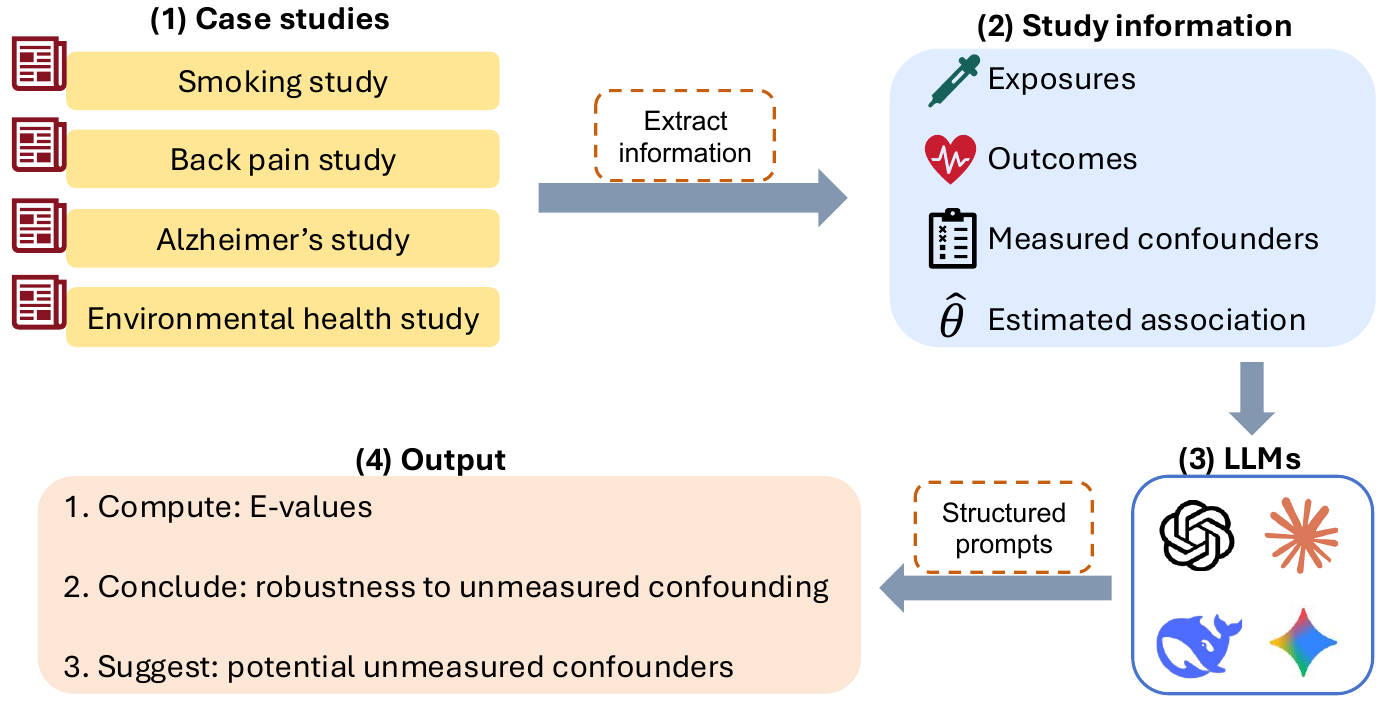}
\caption{Overview of the study workflow. (1) Four published observational studies are used as inputs. (2) Key study information is extracted and (3) provided to large language models (LLMs). (4) The LLMs compute E-values, assess robustness to unmeasured confounding, and suggest potential unmeasured confounders.}
\label{fig:case-study}
\end{figure}

Connected to the idea of Cornfield inequality, \cite{vanderweele2017sensitivity} proposed the E-value as an assumption-lean metric for sensitivity analysis. The E-value represents the minimum association that an unmeasured confounder would need to have with both the exposure and the outcome, such that this unmeasured confounder can fully explain away the observed exposure–outcome effect. The core formula for calculating the E-value is
\begin{equation*}
  E\text{-value} = RR + \sqrt{RR \cdot \left| RR - 1 \right|},
\end{equation*}
where ``$RR$" represents the risk ratio of exposure to outcome in observational studies after adjusting for measured covariates. Besides the risk ratio (RR), the E-value also applies to other effect measures, including hazard ratio (HR) and odds ratio (OR). When an observed $RR=3.6$, the formula calculates the E-value as $3.6+\sqrt{3.6×(3.6-1)}=6.66$. Therefore, to explain away an $RR=3.6$,  an unmeasured confounder would need to be associated with both exposure and outcome by at least 6.66-fold risk ratio each. A larger E-value suggests that it is less plausible that unmeasured confounding could explain away the estimated association, and vice versa.

Figure \ref{fig:case-study} illustrates the overall workflow of our study. We select four published observational studies. From each study, we extract key information, including the exposure, outcome, measured confounders, and the estimated associations/effects. These study-specific inputs are then provided to LLMs via structured prompts. Finally, the LLMs generate the outputs of interest: calculating E-values, interpreting the robustness of the observed association to unmeasured confounding, and suggesting plausible unmeasured confounders.

\subsection{Sources of case studies}

To evaluate the performance of LLMs in sensitivity analysis, we select four published observational studies from different fields: a smoking study, a back pain study, an Alzheimer's study, and an environmental health study. We briefly summarize these four studies as follows:
\par

\begin{itemize}
    \item Smoking study \citep{bellou2021tobacco}: The study investigated whether smoking-related exposures affected the hazard of idiopathic pulmonary fibrosis. 
    \item Back pain study \citep{ikeda2023changes}: The study investigated  whether changes in body mass index affected the risk of back pain.
    \item Alzheimer's study \citep{yaghmaei2024combined}: The study investigated whether Alzheimer's treatment affected the five-year survival probability of Alzheimer’s disease patients. 
    \item Environmental health study \citep{raffetti2018polychlorinated}: The study investigated whether changes in serum polychlorinated biphenyls (PCB), which are toxic, bioaccumulative environmental pollutants, affected the risk of hypertension.
\end{itemize}

Table \ref{tab: case study} summarizes more detailed characteristics from these four studies. All four studies clearly documented the exposures, outcomes, measured confounders, and estimated effect sizes. In addition, they all cited \citet{vanderweele2017sensitivity}, and calculated and reported the E-values. Since some studies included multiple exposures, a total of 11 E-values were calculated across the four studies. Except for the Alzheimer's study, the other three studies explicitly made conclusions and interpretations regarding the level of unmeasured confounding based on the calculated E-values.

\begin{table}[H]
\centering
\centerline{\includegraphics[width = \textwidth]{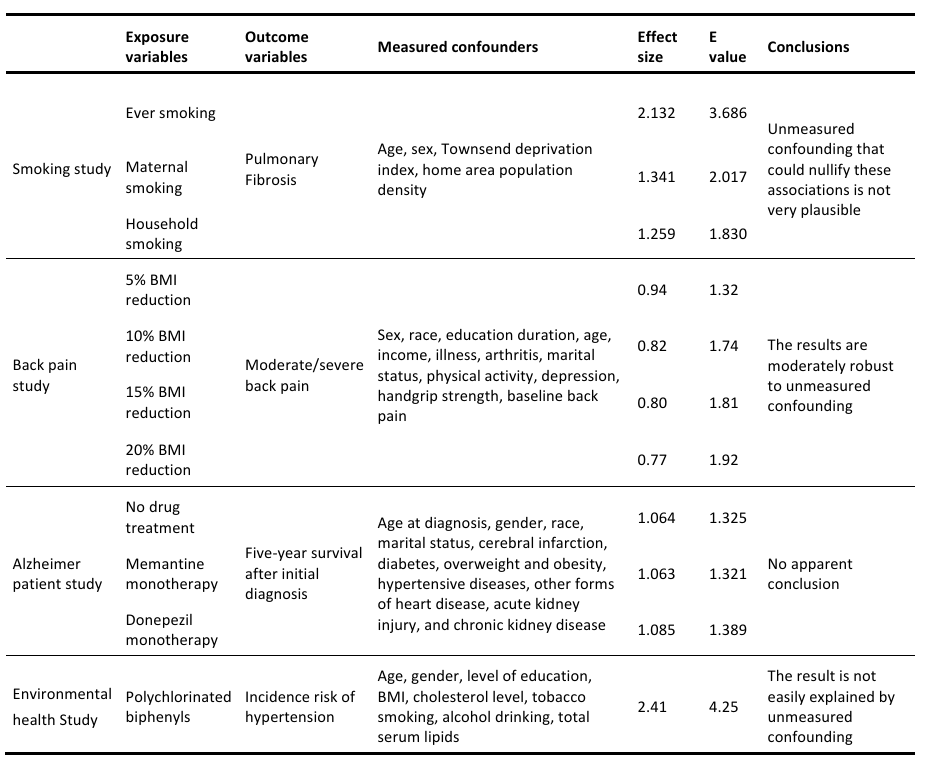}}

\caption{Summary characteristics of the four observational studies. The exposures, outcomes, measured confounders, and effect sizes in those paper will be used as the input to the large language models. BMI:  body mass index. } 
\label{tab: case study}
\end{table}

\subsection{Analysis to assess the performance of LLM in sensitivity analysis}

Our analysis aims to assess the LLMs' ability to conduct sensitivity analysis based on summarized information from an observational study. We provide the LLMs with key information from each of four case studies:  exposures, outcomes, measured confounders, and the estimated associations  (Table \ref{tab: case study}). The E-values and conclusions reported in those studies will serve as the ground truth for evaluating the LLMs' performance.

We evaluate four popular LLMs: ChatGPT-5 mini \citep{achiam2023gpt},  Claude 4.5 Opus \citep{claude45opus2025}, DeepSeek V3 \citep{liu2024deepseek}, and Gemini-2.5-pro \citep{team2023gemini}. We develop a thorough prompt strategy to guide those LLMs for sensitivity analysis. The full prompts are included in Section S1 of the supplementary file, which is available in the GitHub repository of the authors. Our prompt method begins by using an initialization prompt to provide the LLMs with relevant context:

\texttt{You are a helpful epidemiologist and causal inference expert with a clinical background, specializing in assisting researchers with sensitivity analysis. You are knowledgeable in both the calculation and interpretation of Cornfield inequalities and E-values, and you have a deep understanding of their theories and clinical implications.}

We then provide detailed task prompts. First, as a quantitative evaluation, we let LLMs calculate E-values based on the information provided from each input study. The explicit formula of calculating E-values is not provided, as to assess if LLMs can independently perform such calculation. The prompt for this task is:

\texttt{1. Calculate the E-value using the appropriate formula}

Then, as a qualitative evaluation, we let the LLMs draw conclusions on how likely unmeasured confounding can explain away the observed association. The prompts are designed to have several steps. Specifically, the LLMs are first instructed to make separate assessments based on Cornfield inequality and the calculated E-value. After considering both perspectives, LLMs are then asked to provide a final conclusion with reasoning. The prompts for those tasks are:

\texttt{2. Evaluate from Cornfield inequality perspective: Consider (a) whether any single unmeasured confounder could possibly have the required strength of association with both exposure and outcome, (b) plausibility of such confounders in this specific context, (c) if any known strong confounders in this context have already been measured. Provide your analysis (1-2 sentences)}

\texttt{3. Evaluate from E-value perspective: Consider (a) the magnitude of the calculated E-value, (b) whether an unmeasured confounder with such strength is plausible given the exposure-outcome relationship. Provide your analysis (1-2 sentences)}

\texttt{4. Please consider BOTH Cornfield inequality and E-value evaluations above, and draw a conclusion: conclude whether unmeasured confounding is "unlikely", "possibly", or "highly likely" to explain away the observed association. Provide a comprehensive reason (2-3 sentences) that synthesizes both perspectives}

Finally, as an exploratory evaluation, we let the LLMs suggest potential unmeasured confounders relevant to the input study. This task is intended to assess whether LLMs could recognize possible sources of confounding beyond the variables reported in the original studies. Such suggestions can help researchers better identify important variables to measure in the design of their studies.

\texttt{5. Identify 3 potential unmeasured confounding variables relevant to this specific exposure-outcome relationship}

\subsection{Implementation}

As proprietary LLMs cannot be deployed locally, all inference is conducted through the APIs of the LLMs. To optimize accuracy and reproducibility, we set the temperature to 0 to reduce output randomness and suppress hallucinated content. Meanwhile, to accommodate long text outputs containing all reasoning steps, we set the maximum number of generated tokens to 2000. All inputs and outputs are processed  through Python scripts to ensure consistency in format, token budget, and output structure across models. The codes are included in a GitHub repository \url{https://github.com/qingyan16/LLMs-for-sensitivity-analysis}.

\section{Results}

\subsection{LLMs' quantitative ability in calculating E-values}


\begin{table}[htbp]

\centering
\small
\setlength{\tabcolsep}{4pt}

\renewcommand{\arraystretch}{1.25}

\begin{adjustbox}{max width=\textwidth}
\begin{tabular}{@{}l l l c c c c c c@{}}
\toprule
\textbf{Study} 
& \textbf{\makecell[l]{Exposure\\Variables}} 
& \textbf{\makecell[l]{Outcome\\Variables}} 
& \textbf{\makecell{Effect\\Size}} 
& \textbf{\makecell{True E\\value}} 
& \multicolumn{4}{c}{\textbf{Bias of LLM-calculated E-value}} \\
\cmidrule(lr){6-9}
& & & & 
& \textbf{\makecell{ChatGPT}} 
& \textbf{\makecell{Claude}} 
& \textbf{\makecell{DeepSeek}} 
& \textbf{\makecell{Gemini}} \\
\midrule

\multirow{3}{*}{Smoking Study}
& Ever smoking
& \multirow{3}{*}{Pulmonary Fibrosis}
& 2.132 & 3.686 & 0.0 & 0.0 & 0.23 & 0.0 \\
& Maternal smoking
& 
& 1.341 & 2.017 & 0.0 & 0.0 & -0.01 & 0.0 \\
& Household smoking
& 
& 1.259 & 1.830 & 0.0 & 0.0 & 0.10 & 0.0 \\

\addlinespace[0.6em]

\multirow{4}{*}{Back Pain Study}
& 5\% BMI reduction
& \multirow{4}{*}{\makecell[l]{Moderate/severe\\back pain}}
& 0.94 & 1.32 & 0.0 & 0.0 & 0.12 & 0.0 \\
& 10\% BMI reduction
& 
& 0.82 & 1.74 & 0.0 & 0.0 & 0.10 & 0.0 \\
& 15\% BMI reduction
& 
& 0.80 & 1.81 & 0.0 & 0.0 & 0.14 & 0.0 \\
& 20\% BMI reduction
& 
& 0.77 & 1.92 & 0.0 & 0.0 & 0.14 & 0.0 \\

\addlinespace[0.6em]

\multirow{3}{*}{\makecell[l]{Alzheimer\\patient Study}}
& No drug treatment
& \multirow{3}{*}{\makecell[l]{Five-year survival\\after initial diagnosis}}
& 1.064 & 1.325 & 0.0 & 0.0 & 0.10 & 0.0 \\
& Memantine monotherapy
& 
& 1.063 & 1.321 & 0.0 & 0.0 & 0.13 & 0.0 \\
& Donepezil monotherapy
& 
& 1.085 & 1.389 & 0.0 & 0.0 & 0.19 & 0.0 \\

\addlinespace[0.6em]

Environmental health study
& Polychlorinated biphenyls
& Incidence risk of hypertension
& 2.41 & 4.25 & 0.0 & 0.0 & 0.01 & 0.0 \\

\bottomrule
\end{tabular}
\end{adjustbox}
\caption{The E-values calculated by LLMs based on the extracted information of four case studies. The true E-values are obtained from the original observational studies. The bias is the difference between the calculated E-values and the actual reported values. \\ BMI:Body Mass Index.}
\label{tab: Evalues}
\end{table}

This subsection evaluates the quantitative ability of LLMs in calculating the E-values (Table \ref{tab: Evalues}) based on the summarized study information. Since some studies included multiple exposure-outcome pairs, a total of 11 E-value calculations are generated. The calculated E-values from ChatGPT, Claude, and Gemini exactly match the ground truth E-values reported in the original studies, showing zero bias across all cases. However, DeepSeek shows biases ranging from -0.01 to 0.23 relative to the true E-values. Notably, although the explicit formula for calculating the E-value is not provided in the prompts, ChatGPT, Claude, and Gemin can still accurately calculate E-values using  summarized information of  exposures, outcomes, measured confounders, and effect estimates.

\subsection{LLMs' conclusions on robustness to unmeasured confounding based on both Cornfield inequality and the E-value}

\begin{table}[!htbp]

\centering
\small
\setlength{\tabcolsep}{3pt}
\renewcommand{\arraystretch}{1.25}
\begin{adjustbox}{max width=\textwidth}
\begin{tabular}{@{}l l l c c c c c c@{}}
\toprule
\textbf{Study} 
& \textbf{\makecell[l]{Exposure\\Variables}} 
& \textbf{\makecell[l]{Outcome\\Variables}} 
& \textbf{\makecell{Effect\\Size}} 
& \textbf{\makecell{E\\value}} 
& \multicolumn{4}{c}{\textbf{Conclusions generated from LLMs}} \\
\cmidrule(lr){6-9}
& & & & 
& \textbf{\makecell{ChatGPT}} 
& \textbf{\makecell{Claude}} 
& \textbf{\makecell{DeepSeek}} 
& \textbf{\makecell{Gemini}} \\
\midrule

\multirow{3}{*}{Smoking study}
& Ever smoking 
& \multirow{3}{*}{Pulmonary Fibrosis}
& 2.132 & 3.686 & Unlikely & Unlikely & Possibly & Possibly \\
& Maternal smoking 
& 
& 1.341 & 2.017 & Possibly & Possibly & Possibly & Possibly \\
& Household smoking 
& 
& 1.259 & 1.830 & Possibly & Possibly & Possibly & Possibly \\

\addlinespace[0.6em]

\multirow{4}{*}{Back pain study}
& 5\% BMI reduction 
& \multirow{4}{*}{\makecell[l]{Moderate/severe\\back pain}}
& 0.94 & 1.32 & Possibly & Highly likely & Possibly & Possibly \\
& 10\% BMI reduction 
& 
& 0.82 & 1.74 & Possibly & Possibly & Possibly & Possibly \\
& 15\% BMI reduction 
& 
& 0.8 & 1.81 & Possibly & Possibly & Possibly & Possibly \\
& 20\% BMI reduction 
& 
& 0.77 & 1.92 & Possibly & Possibly & Possibly & Possibly \\

\addlinespace[0.6em]

\multirow{3}{*}{\makecell[l]{Alzheimer\\patient study}}
& No drug treatment
& \multirow{3}{*}{\makecell[l]{Five-year survival\\after initial\\diagnosis}}
& 1.064 & 1.325 & Highly likely & Highly likely & Possibly & Highly likely \\
& Memantine monotherapy
& 
& 1.063 & 1.321 & Highly likely & Highly likely & Possibly & Highly likely \\
& Donepezil monotherapy
& 
& 1.085 & 1.389 & Possibly & Highly likely & Possibly & Highly likely \\

\addlinespace[0.6em]

Environmental health study
& Polychlorinated biphenyls
& Incidence risk of hypertension
& 2.41 & 4.25 & Unlikely & Unlikely & Possibly & Unlikely \\

\bottomrule
\end{tabular}
\end{adjustbox}
\caption{Conclusions generated by LLMs based on perspectives of both Cornfield inequality and E-value. Unlikely: the relationship between exposures and outcome variables is \textit{unlikely} to be explained by unmeasured confounding. The interpretations for `Possibly' and `Highly likely' follow similarly.}
\label{conclusions}
\end{table}

Table \ref{conclusions} shows the qualitative conclusions generated by the LLMs based on both the Cornfield inequality and the E-value. Detailed reasoning outputs from both perspectives are provided in Supplementary File Section S2. From the reasoning outputs, the LLMs demonstrate the ability to apply the Cornfield inequality properly. For example, in the smoking study,  the LLMs translate every observed effect size into the required strength of association that an unmeasured confounder would need to have with both the exposure and the outcome. In particular, the reasoning from ChatGPT and Claude varies with the magnitude of the effect size. For a larger effect (RR = 2.132) in the smoking study, both models respond that it is ``less plausible" of a single unmeasured confounder explaining away the association, while for smaller effects (e.g., RR = 1.259), both models conclude such unmeasured confounding is plausible.

Across all cases in Table \ref{conclusions}, GPT, Claude, and Gemini generate final conclusions that differentiate appropriately, ranging from  ``unlikely" to ``possibly" to ``highly likely". Their conclusions largely align with the magnitude of the effect sizes and the calculated E-values. In contrast, DeepSeek consistently concludes that it is ``possible"  that unmeasured confounding can explain away the observed effect, regardless of effect size or calculated E-value. Therefore, we will focus on the conclusions generated by ChatGPT, Claude, and Gemini in the following paragraphs.

For cases with intermediate E-values,  GPT, Claude, and Gemini largely agreed, concluding that it is ``possible" that unmeasured confounding could explain away the observed effect, For cases with large E-values, for example, the environmental health study with an E-value of 4.25,  these three models also agree on a conclusion of ``unlikely". However, at lower E-values (less than 1.4), differences emerge in the LLMs' conclusions. For example, in the back pain study with an E-value of 1.32, Claude concludes that unmeasured confounding is ``highly likely" to explain away the observed effect. In contrast, GPT still concludes this case as ``possibly". Such pattern is also observed for the case in the Alzheimer's study with the E-value of 1.389, where Claude again concludes ``highly likely" while GPT concludes ``possibly". These results suggest that when the E-value is small, Claude more readily concludes the observed effect as highly susceptible to unmeasured confounding, while ChatGPT prefers to not conclude the case as ``highly likely".

\subsection{Potential unmeasured confounders suggested by the LLMs. Each LLM is required to suggest top 3 unmeasured confounders.}

Table \ref{tab:llm_confounders}  shows the results of unmeasured confounders suggested by LLMs. Based on the summarized study information, all four models suggest potential unmeasured confounders that are  biologically or epidemiologically reasonable. For example, in the smoking study, all models suggest occupational exposures (e.g. dust) and genetic susceptibility, which are  not measured in the original study but are also risk factors for smoking and pulmonary fibrosis \citep{fujishiro2012occupational, park2021occupational}. 
\par

\begin{table}[!htbp]
\centering
\footnotesize 
\renewcommand{\arraystretch}{1.25}

\begin{adjustbox}{max width=\textwidth}
\begin{tabular}{@{}lllllll@{}}
\toprule
\textbf{Study} & \textbf{Exposures} & \textbf{Outcomes} & \multicolumn{4}{c}{\textbf{Unmeasured confounders suggested from LLMs}} \\ 
\cmidrule(lr){4-7}
& & & \textbf{ChatGPT} & \textbf{Claude} & \textbf{DeepSeek} & \textbf{Gemini} \\ 
\midrule

\makecell[l]{Smoking\\study} & \makecell[l]{Smoking\\related\\exposure} & \makecell[l]{Pulmonary\\Fibrosis} & 
\makecell[l]{1. Genetic predisposition\\2. Occupational exposures\\3. Environmental pollution/\\tobacco exposure} & 
\makecell[l]{1. Occupational and\\Environmental Exposure\\2. Genetic-Related Factors\\3. Chronic Respiratory\\Disease and Infection\\Factors} & 
\makecell[l]{1. Occupational exposure\\to dust/fibers or hazards\\2. Environmental exposure\\(secondhand smoke, indoor\\air pollution)\\3. Genetic predisposition\\to pulmonary fibrosis} & 
\makecell[l]{1. Occupational exposures\\(dust, fumes, chemicals,\\asbestos, silica)\\2. Genetic predisposition\\to pulmonary fibrosis\\3. History of respiratory\\infection} \\ 
\midrule

\makecell[l]{Back pain\\study} & \makecell[l]{BMI\\reduction} & \makecell[l]{Moderate/severe\\back pain} & 
\makecell[l]{1. Occupational physical\\demands\\2. Sleep quality and\\duration\\3. Use of pain medication} & 
\makecell[l]{1. Occupational factors\\2. Sleep quality and\\duration\\3. Genetic predisposition\\to musculoskeletal\\conditions} & 
\makecell[l]{1. Occupational physical\\demands\\2. Psychological/\\psychosocial stress\\3. Genetic predisposition\\to chronic pain, sleep\\quality or prior spinal\\injuries} & 
\makecell[l]{1. Occupational factors\\(physical demands,\\job type)\\2. Sleep quality and\\duration\\3. Smoking status, pain\\medication use or prior\\back injuries} \\ 
\midrule

\makecell[l]{Alzheimer\\patient\\study} & \makecell[l]{Alzheimer\\treatments} & \makecell[l]{Five-year\\survival after\\initial\\diagnosis} & 
\makecell[l]{1. Dietary habits\\2. Sleep quality\\3. Stress levels} & 
\makecell[l]{1. Socioeconomic status\\and access to healthcare\\2. Disease severity or\\stage (including baseline\\status)\\3. Lifestyle factors or\\concurrent medications} & 
\makecell[l]{1. Socioeconomic status\\2. Access to healthcare\\3. Disease severity at\\diagnosis or baseline} & 
\makecell[l]{1. Socioeconomic status\\or lifestyle factors\\2. Comorbidities and\\disease severity\\3. Adherence to treatment\\(Memantine, Donepezil)} \\ 
\midrule

\makecell[l]{Environmental\\health\\study} & \makecell[l]{Polychlorinated\\biphenyls} & \makecell[l]{Incidence risk\\of hypertension} & 
\makecell[l]{1. Dietary factors\\2. Physical activity\\3. Environmental stressors} & 
\makecell[l]{1. Occupational co-exposures\\(other industrial chemicals,\\heavy metals)\\2. Dietary patterns (fish\\consumption frequency\\and source)\\3. Physical activity level\\and fitness} & 
\makecell[l]{1. Physical activity level\\2. Dietary patterns (e.g.,\\sodium intake, fruit/\\vegetable consumption)\\3. Occupational exposure\\to other environmental\\toxins} & 
\makecell[l]{1. Genetic predisposition\\to hypertension\\2. Dietary intake of\\sodium and potassium\\3. Occupational exposure\\to other environmental\\toxins} \\ 
\bottomrule
\end{tabular}
\end{adjustbox}
\caption{Unmeasured confounders suggested by large language models across four  observational studies.}
\label{tab:llm_confounders}
\end{table}

There are differences in the specificity and focus of the suggested confounders among different models. In general, Claude, DeepSeek, and Gemini tend to provide more specific suggestions. In the environmental health study, Claude, DeepSeek, and Gemini all suggest ``dietary patterns" as a potential unmeasured confounder; however, Claude specifically mentions ``fish consumption"; DeepSeek mentions ``fruit/vegetable consumption"; Gemini mentions ``Dietary intake of sodium and potassium".  In contrast, ChatGPT’s suggestions were relatively broader across all four studies, often using general terms without providing more details or examples.

\section{Discussion}


Our work investigates the performance of four widely used large language models (ChatGPT, Claude, DeepSeek, and Gemini) in conducting sensitivity analyses. Using studies from four different clinical and epidemiological domains, we provide the LLMs with summarized study information including the exposures, outcomes, measured confounders, and effect estimates. We develop structured prompts to guide the LLMs to perform different tasks for sensitivity analysis. Overall, the results show that ChatGPT, Claude, and Gemini can accurately calculate E-values,  properly assess robustness to unmeasured confounding, and identify reasonable potential unmeasured confounders.

The conclusions generated by the LLMs regarding how likely unmeasured confounding could explain away the observed associations are appropriately differentiated. In contrast, the original observational studies typically provide a single qualitative interpretation, even when multiple exposure-outcome cases are reported. For example, in the smoking study, although the reported effect sizes range from 1.259 to 2.132, this article drew a single conclusion that it is not plausible for unmeasured confounding to explain away the observed effects. However, the ChatGPT conclusions distinguish among exposure-outcome pairs, suggesting that unmeasured confounding is “unlikely” when the effect size is 2.132 but “possibly” when the effect size is 1.259. These conclusions, which are based on both the Cornfield inequality and the E-value, suggest that LLMs can generate interpretations that align with the magnitude of effect estimates and corresponding E-values.

As this is a case study, we include only four published studies as inputs to the LLMs. In future research, we could use a larger number of observational studies that apply the Cornfield inequality and the E-value, allowing for a more thorough evaluation.

LLMs are increasingly used in the research community \citep{kobak2024delving, singhal2025toward}, and this study fills an important gap by assessing  their ability to support sensitivity analysis. We hope that the prompt strategies developed in this work (Supplementary File Section S1) can serve as useful tools to help researchers better assess the susceptibility to unmeasured confounding based on both the Cornfield inequality and the E-value.

\section*{Acknowledgements}
The authors would like to thank the journal \textit{Observational Studies} for their announcement of the special issue on Cornfield inequality and sensitivity analyses, which motivated this research.

\label{app:theorem}





\vskip 0.2in
\bibliography{sample}

\end{document}